\documentstyle[twoside,fleqn,espcrc2,epsf]{article}

\newcommand{\AmS}{{\protect\the\textfont2
  A\kern-.1667em\lower.5ex\hbox{M}\kern-.125emS}}

\title{The Nature  of  Singularity  in  Bianchi  I Cosmological
       String  Gravity   Model   with   Second  Order  Curvature
       Corrections}

\author{S.Alexeyev, A.Toporensky, V.Ustiansky \\
    {\it Sternberg Astronomical Institute, Moscow State University,
         Universitetsky Prospekt, 13, \\ Moscow 119899, Russia}}

\begin{document}

\begin{abstract}
We investigate Bianchi I cosmological model in the theory of a dilaton
field coupled to  gravity  through a  Gauss-Bonnet  term. Two type  of
cosmological singularity are distinguished. The former is analogous to
the Einstein gravity singularity, the latter (which does not appear in
classical General Relativity)  occurs when the main determinant of the
system  of  field equations vanishes. An analogy  between  the  latter
cosmological singularity and  the singularity inside a black hole with
a dilatonic hair is discussed.  Initial  conditions,  leading to these
two  types  of  cosmological  singularity  are   found  via  numerical
integration of the equation of motion.
\vspace{1pc}
\end{abstract}

\maketitle

General Relativity (GR) as leading theory of gravitational interaction
at the classical level  demonstrates  a lot of interesting properties,
for example, the  existence of space time singularities \cite{e1}. The
main ``negative'' feature of GR is that it is not  renormalisable, so,
it can not be quantized  by  the usual quantum field theory  methods.
The possible way to construct the  quantum theory of gravity is to use
string/M theory  as the most promising  candidate for the  ``theory of
all   physical   interactions''.  After   compactification   to   four
dimensional space time one obtains some effective gravitational theory
which includes GR as  the  zeroth perturbation. This approach contains
more wide set of solutions, and, hence, in addition to the singularity
classes introduced by GR can include new types of singular behavior.

The simplest  version  of  effective  four  dimensional string gravity
action with  the higher order curvature  corrections up to  the second
order  contains  a  coupling  of  the  dilaton with  gravity  via  the
Gauss-Bonnet term as follows \cite{action}
 (we  use the units where $m_{PL}/\sqrt{8
\pi}=1$):
\begin{eqnarray}\label{a0}
S & = & \int d^4 x \sqrt{-g} \biggl[ - \frac12R
+ \frac12\partial_\mu \phi \partial^\mu \phi \nonumber \\
& + & \frac\lambda{16} e^{-2 \phi} S_{GB}  \biggr] ,
\end{eqnarray}
where $R$ is Ricci scalar, $\phi$ is dilaton,  $\lambda$ is dilatonic
string  coupling  constant which is positive. Second order  curvature
correction $\lambda e^{-2\phi} S_{GB}$ represents the  product of the
coupling functions $e^{-2 \phi}$ and Gauss-Bonnet combination $S_{GB}
= R_{ijkl} R^{ijkl} -  4 R_{ij} R^{ij} + R^2$. The solutions  of this
model have extensively  been  studied in  the  literature, both in  a
perturbative \cite{e3} and numerical \cite{e4,e5} approaches.

One  important  aspect   is  that  when  the  second  order  curvature
correction is  taken into account a  wide class of  solutions contains
new types of singularities (earlier it  was studied in $d \ge 4$ space
time case, see, for  instance,  \cite{e6} and references therein). For
example,  during  the   investigation   of  the  black  hole  solution
(asymptotically flat, spherically  symmetric and static) a new type of
black hole  inner singularity was found. It occurs  at the finite (non
zero) radius  inside black hole and has the  topology $S^2 \times R^1$
(an infinite tube in  time  direction). This is curvature singularity,
though it is week in Tipler's terminology \cite{Tipler}.  If one works
in curvature gauge
\begin{eqnarray}\label{e01}
ds^2 = \Delta dt^2 - \frac{\sigma^2 }{\Delta } dr^2 - r^2
(d \theta^2 + \sin^2 \theta d \varphi^2) \nonumber,
\end{eqnarray}
where $\Delta = \Delta(r)$ and $\sigma = \sigma(r)$, Einstein-Lagrange
equations can be written in a matrix form
\begin{eqnarray}\label{e02}
a_{i1} \Delta'' + a_{i2} \sigma' + a_{i3} \phi'' = b_i,
\end{eqnarray}
where $i=1,2,3$, $a_{ij}=a_{ij}(\Delta, \Delta', \sigma, \phi, \phi')$
and $b_i = b_i (\Delta, \Delta', \sigma, \phi, \phi')$.

The  describing  singularity   (we   will  call  it  as  ``determinant
singularity'') occurs when  the second part (in quadratic brackets) of
the main system determinant
\begin{eqnarray} \label{e03}
D_{main} = \Delta \biggl[ A \Delta^2 + B \Delta + C \biggr] ,
\end{eqnarray}
($A$,  $B$  and  $C$  depend  upon  $\Delta',  \sigma,  \phi,  \phi'$)
vanishes.  The  asymptotic  behavior  of  the   metric  and  dilatonic
functions near this singularity $r_s$ is \cite{e5}
\begin{eqnarray}\label{e03a}
\Delta & = & d_s + d_1 (\sqrt{r - r_s})^2
+ d_2 (\sqrt{r - r_s})^3 + \ldots,
\nonumber \\
\sigma & = & \sigma_s + \sigma_1 \sqrt{r - r_s}
+ \sigma_2 (\sqrt{r - r_s})^2 + \ldots,\\
\phi & = & \phi_s + \phi_1 (\sqrt{r - r_s})^2
+ \phi_2 (\sqrt{r - r_s})^3 + \ldots,\nonumber
\end{eqnarray}
where $d_i$,  $\sigma_i$, $\phi_i$ are  numerical factors and \\
$r - r_s \ll 1$.

\begin{figure}[hb]
\epsfxsize=\hsize
\centerline{\epsfbox{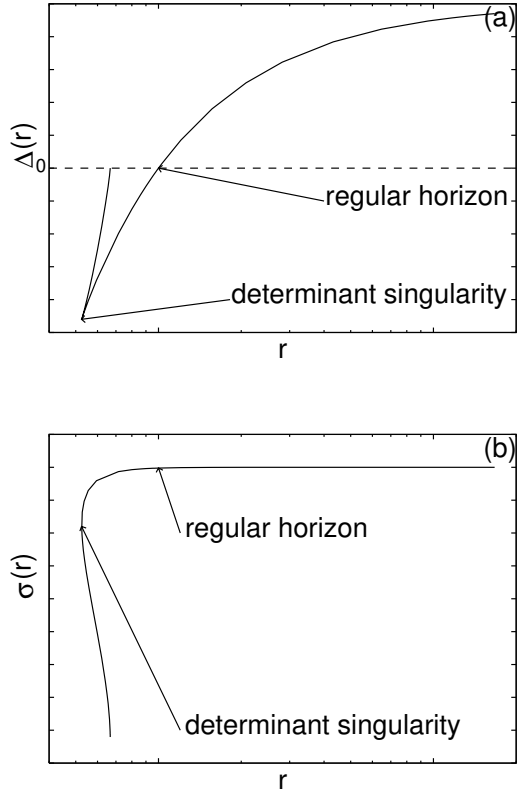}}
\caption{The  behavior  of the metric functions $\Delta$ (presents  as
$\partial^2 \Delta / \partial r^2$  in  field  equations) and $\sigma$
(presents  as  $\partial  \sigma  /  \partial  r$ in field  equations)
against radial coordinate $r$ in black hole case when the second order
curvature corrections are taken into account}
\label{blackhole}
\end{figure}

A numerical example is shown on  Fig.\,\ref{blackhole}(a,b) taken from
\cite{e5}. The  solution was obtained with  the help of  the numerical
method based on the integration over an additional parameter along the
solution trajectory described in Ref. \cite{e5}. Some new mathematical
aspects of this  strategy were additionally studied in Ref \cite{e51}.
The  solution  consists  of  two branches which merge  at  determinant
singularity  $r_s$  and  solution  could not be continued  further  in
radial coordinate. It is necessary  to  point out that all the  metric
functions  (independently  of their maximal derivative order) work  as
const$/\sqrt{r -  r_s}$  in  Riemannian  tensor,  therefore, cutvature
square $I=R_{ijkl}R^{ijkl}$ diverges.

The  main  goal  of  this paper  is to  indicate  that  this  kind of
singularity is  also  significant  in  cosmology.  Previously, it was
found  in  rather  complicated  models  such  as those elaborated  to
describe a dynamical compactification (we think that it  is the first
appearance of this kind  of  singularity in cosmology) \cite{dgp},
in multidimensional Lovelock gravity \cite{Kitaura-Wheeler}  and
in  Bianchi  I  \cite{Kawai:1999bn,Yajima:2000gk}   and   Bianchi  IX
\cite{Yajima:2000gk} with moduli fields. In our paper  we show that
the determinant singularity occurs  in  the  Bianchi I cosmology
(the line element is
\begin{eqnarray}\label{e04}
ds^2 = dt^2 - a^2(t) d x_1^2 - b^2(t) d x_2^2 - c^2(t) d x_3^2 ,
\end{eqnarray}
$a(t)$, $b(t)$, $c(t)$ are the  scale  factors) in the theory with the
simplest stringy motivated second order curvature correction (1).

Introducing three  Hubble  parameters $p(t)$, $q(t)$, $r(t)$ according
to following definition
$$
p(t)=\frac{\dot a}a,\quad q(t)=\frac{\dot b}b,\quad r(t)=\frac{\dot c}c,
$$
where the dot denotes $d\,/d\,t$, the  variation of action
(\ref{a0}) with respect to the metric components and the dilaton field
gives the following equations of motion:
{\mathindent=0pt
\begin{eqnarray}
& pq+qr+rp+24\dot fpqr-\dot\phi^2/2 = 0, \label{eqv-constr}\\
&(1+8r\dot f)(\dot q+q^2) + (1+8q\dot f)(\dot r+r^2)
\nonumber \\
& \kern1.5cm {}+ (1+8\ddot f)qr+\dot\phi^2/2=0,\\
&(1+8r\dot f)(\dot p+p^2) + (1+8p\dot f)(\dot r+r^2)
\nonumber \\
& \kern1.5cm {}+ (1+8\ddot f)rp+\dot\phi^2/2=0,\\
&(1+8q\dot f)(\dot p+p^2) + (1+8p\dot f)(\dot q+q^2)
\nonumber \\
& \kern1.5cm {} + (1+8\ddot f)pq+\dot\phi^2/2=0,\\
&\ddot\phi=-(p+q+r)\dot\phi +8f'(\dot pqr+p\dot qr
\nonumber \\
& \kern1.5cm {} + pq\dot r+pqr(p+q+r)),\label{eqv-phi}
\end{eqnarray}}
where   $f(t)=\frac\lambda{16}e^{-2\phi}$,   and  the   prime  denotes
differentiation  with   respect   to   $\phi$.   For   mostly  obvious
presentation  of  the  numerical  results  one  have to introduce  new
variables $h(t)$, $\alpha(t)$, $\beta(t)$
$$
p=h+\alpha+\sqrt3\beta,\quad
q=h+\alpha-\sqrt3\beta,\quad
r=h-2\alpha,
$$
where $h$  is a generalization of a Hubble  parameter of the isotropic
case and  describes the expansion rate  of the universe,  $\alpha$ and
$\beta$ are parameters of anisotropy.

It should be emphasized that the presence of the Gauss-Bonnet term can
in  principle  allow  the  violation  of  the  weak  energy  condition
\cite{e4}.  This means  that  it is possible  for  the spatially  flat
universe to make a transition from expansion to  contraction (see, for
example, \cite{Turner}) what can also be easily seen directly from the
constraint equation  (\ref{eqv-constr}).  Indeed,  in Einstein gravity
the   geometrical   part   of    constraint    (\ref{eqv-constr})   is
$3h^2-3\alpha^2 -3\beta^2$ which  is  nonpositive if $h=0$ and, hence,
this case  is inconsistent with the  constraint equation. Even  if the
Gauss-Bonnet term is  taken  into account  in  the isotropic case  the
correction $24 \dot f pqr$ vanishes at $h=0$, so the  Einstein gravity
result ($h$ can not change it's sign in the spatially flat homogeneous
universe) remains true. However, in Bianchi  I  case  this  correction
contains a  {\it product} of Hubble  parameters which can  be positive
when the {\it sum} of the Hubble parameters  ($h=1/3 (p+q+r)$) becomes
zero  and  the condition $h=0$ no longer  contradicts  the  constraint
(\ref{eqv-constr}).

The equations  of  motion  (\ref{eqv-constr})--(\ref{eqv-phi}) can be
integrated  numerically.  They  have  5  degrees  of  freedom.  Using
constraint (\ref{eqv-constr}) one can reduce this number to  4. As it
was  done  previously \cite{e5} we calculate all dynamical  variables
checking  the  accuracy  with  the  help  of the constraint  equation
(\ref{eqv-constr}). In our calculation the left-hand side 
 of the constraint equation
did  not exceed $10^{-8}$ unless we are very close to an ordinary
singularity. 
  In  all  our  numerical  investigations  we  put
$\lambda = 1$ for simplicity.

The  main  determinant $D_{main}$ is equal to:
{\mathindent=0pt
\begin{eqnarray}
&&D_{main}=3\lambda^4e^{-8\phi}\dot\phi^2r^2p^2q^2\nonumber\\
&&-2\lambda^3e^{-6\phi}\dot\phi(p^2q^2r+p^2qr^2+pq^2r^2)\nonumber\\
&&-2\lambda^3e^{-6\phi}\dot\phi^3rpq \nonumber \\ && -\lambda^2e^{-4\phi}
(p^2q^2+p^2r^2+q^2r^2)\nonumber\\
&&+2\lambda^2e^{-4\phi}(p^2rq+pq^2r+pqr^2)\nonumber\\
&&+2\lambda^2e^{-4\phi}\dot\phi^2(pq+pr+qr) \nonumber \\ &&
-2\lambda e^{-2\phi}\dot\phi(p+q+r)+2\nonumber
\end{eqnarray}}
The  determinant
singularity   occurs   when   $D_{main}$   vanishes.   Our   numerical
investigations  show  that  it can happen  for  a  rather  wide set of
initial data (see below). The asymptotic  form  of  Hubble  parameters
behavior near this singularity $t_s$ is (we checked it numerically and
solved the equations on expansion  coefficients $h_{i\mu}$, there are no
any kinds of contradictions)
\begin{eqnarray}\label{e05}
h_i & = & {h_i}_s + {h_i}_1 \sqrt{t - t_s}
+ {h_i}_2 (\sqrt{t - t_s})^2 + \ldots, \nonumber\\
\phi & = & \phi_s + \phi_1 (\sqrt{r - r_s})^2
+ \phi_2 (\sqrt{r - r_s})^3 + \ldots
\end{eqnarray}
where $h_i = p,q,r$  and $t - t_s \ll 1$.

\begin{figure}[ht]
\epsfxsize=.9\hsize
\centerline{\epsfbox{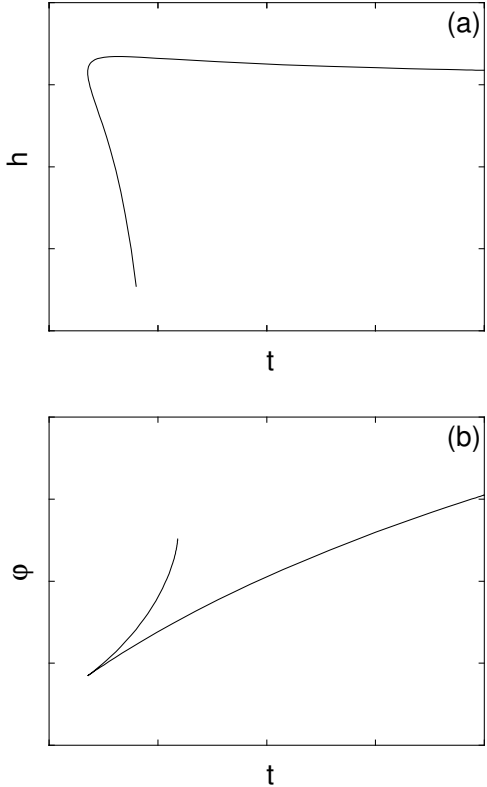}}
\caption{The  behavior  of  the  Hubble functions $h_i$  (presents  as
$\partial h_i  / \partial t$  in field equations) and $\phi$ (presents
as  $\partial^2  \phi  /  \partial  t^2$  in field equations)  against
coordinate  $t$  in Bianchi  I  case when  the  second order  curvature
corrections are taken into account}
\label{bianchii}
\end{figure}
Comparing (11) with (4) we can see that these two expansions are 
absolutely analogous. The expansion properties of functions having
the same maximal derivative order in the equations of motion
(the first order for $\sigma$ and $h_i$, the second order for $\Delta$,
and $\phi$) are identical. It means that the properties of the determinant
cosmological singularity are the same as the properties of the determinant
Schwarzschild black hole singularity (keeping in mind, of course, that
metric functions in cosmological case depend on time while in black hole
 they depend on the radial coordinate).

A  numerical example is  plotted  in  Fig.\,\ref{bianchii}.  Again, we
can  see  two
branches of  solution and we  can not continue these solutions further
in time.

In  our   investigation   we   calculated  two-dimensional  slices  in
$(\alpha,\beta)$-plane of initial condition space with fixed values of
$h$ and $\phi$. The initial  value  of $\dot\phi$ has been taken  from
the constraint equation, which is quadratic with respect to $\dot
\phi$. The results below corresponds to the choise of the largest root 
of the quadratic equation. Each of the slices represent a $100\times100$
grid. For  each node of the grid  we calculated  both future and  past
solutions over the  time  interval of $10^5$ in  our  units. There are
three possible outcomes: ordinary singularity, determinant singularity
and a  nonsingular behavior. Such  slices have been calculated for the
values of $\phi$ in the interval from $\phi=-15$ to $\phi=+15$ and for
the values of $h$ in the interval from $h=1\cdot10^{-5}$ to $h=1$.

\begin{figure}
\epsfxsize=0.7\hsize
\centerline{\epsfbox{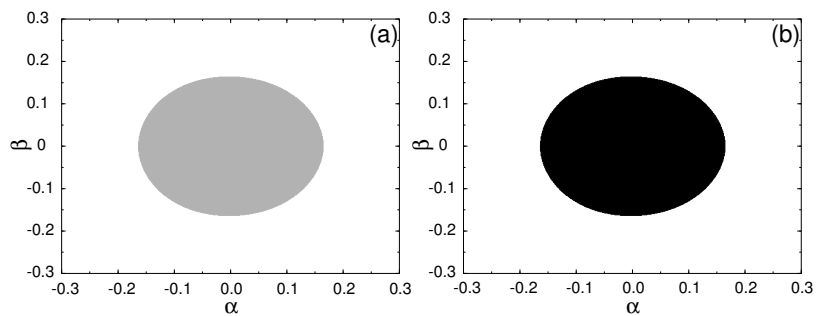}}
\caption{A slices of initial condition space in $(\alpha, \beta)$-plane for
$h=0.16$ and $\phi=0$. Figure (a) corresponds to the past solutions and
figure (b) corresponds to the future solutions.
The  black color  corresponds  to  non-singular  solutions,  the
dark-gray color corresponds to ordinary  singular  solutions  and  the
light-gray   color   corresponds   to   solutions   with   determinant
singularity. The white  color indicates prohibited regions in the initial
condition space.
}
\label{phi0}
\end{figure}

\begin{figure}
\epsfxsize=0.7\hsize
\centerline{\epsfbox{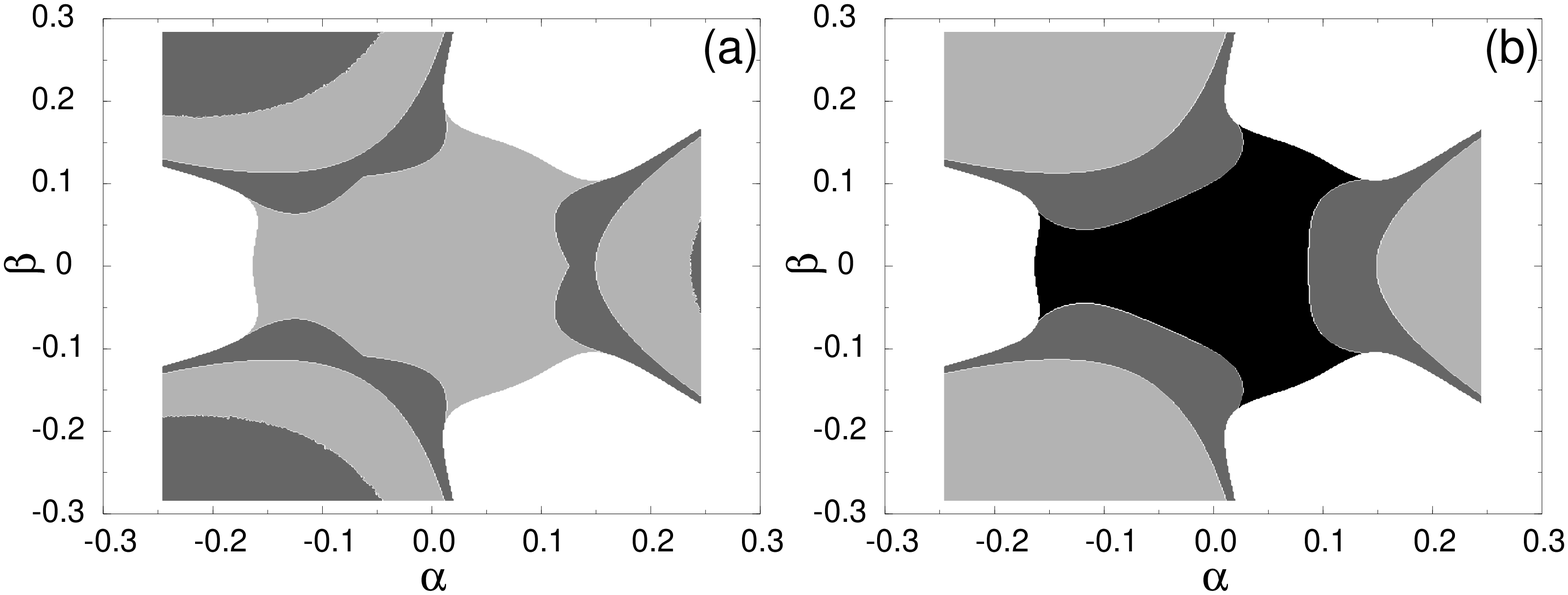}}
\caption{The similar slices for $h=0.16$ and $\phi=-1$.}
\label{phi1}
\end{figure}

\begin{figure}
\epsfxsize=0.7\hsize
\centerline{\epsfbox{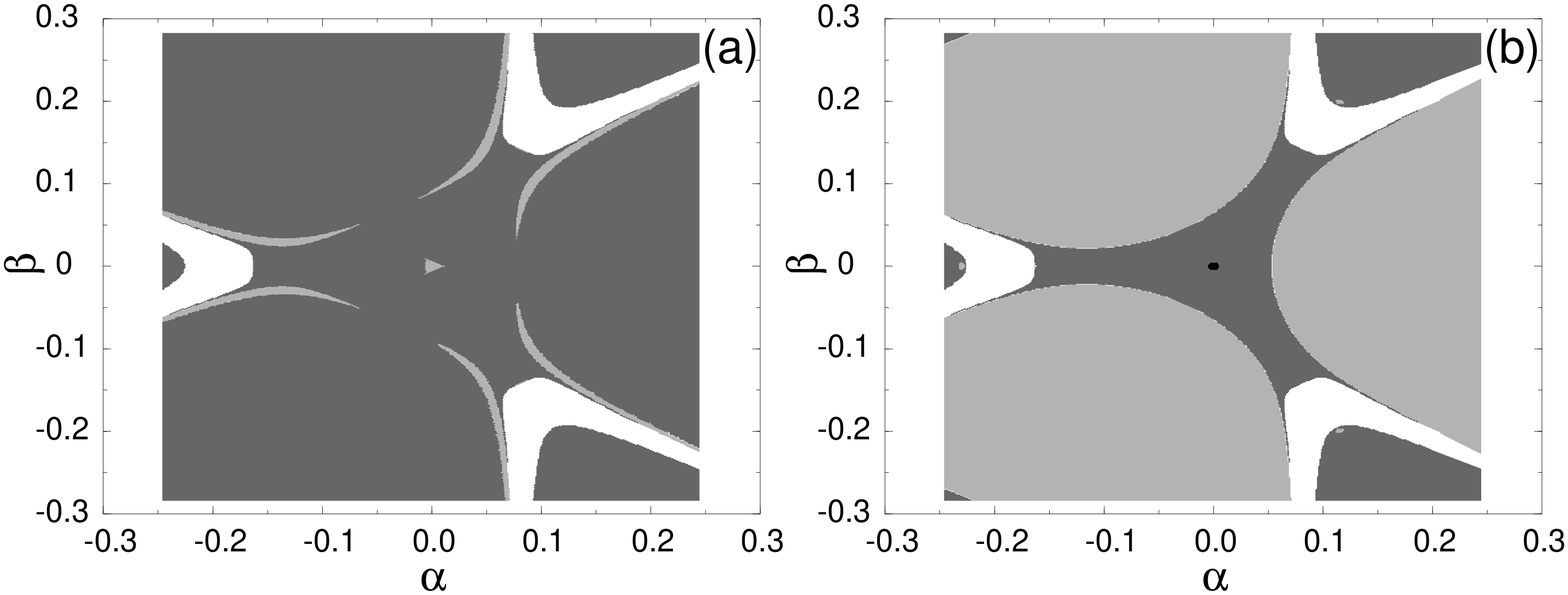}}
\caption{The similar slices for $h=0.16$ and $\phi=-2$.}
\label{phi2}
\end{figure}

Each  of  the numerical solutions has been  analyzed  and  classified.
Finally, for each slice we obtained picture like one shown  in
Fig.\,\ref{phi0}--\ref{phi2}.
Here  the  black color  corresponds  to  non-singular  solutions.  The
dark-gray color corresponds to ordinary  singular  solutions  and  the
light-gray   color   corresponds   to   solutions   with   determinant
singularity. The white  color  indicates prohibited regions in initial
condition space where none of the solutions can start.

For  large  values  of   $\phi$   the  allowed  region  in  $(\alpha,
\beta)$-plane  has a compact subregion of approximately circular shape
with  the  center  in  point $(0,0)$ (see Fig.\,\ref{phi0}).  This  can
be  easily   seen   from constraint   equation
(\ref{eqv-constr})  if  the  term  proportional  to  $e^{-2\phi}$  is
neglected. Numerical calculations show  that  in this case all future
solutions are non-singular while all past solutions are singular with
$D_{main} \to 0$.  There  is only one exception  from  this rule: the
past  solution  for $\alpha=\beta=0$  (the  isotropic  case)  has  an
ordinary singularity.

\begin{figure}
\epsfxsize=.9\hsize
\centerline{\epsfbox{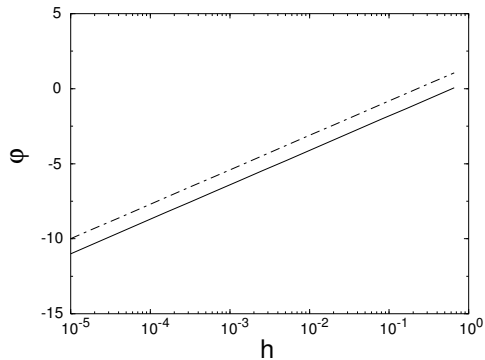}}
\caption{The  minimal  initial values of $\phi$ which allow  solutions
non-singular in future (solid line) and the maximal  initial values of
$\phi$ which allow singular solutions in future (dot-dashed line).}
\label{pminmax}
\end{figure}

With decreasing  of  $\phi$  this  picture  changes dramatically. The
shape of  allowed  region  undergoes  substantial transformation (see
Fig.\,\ref{phi2}): the former compact subregion becomes connected whith
the outer non-compact one. The more  important feature
is  that
some  trajectories meet  a  singularity in future,  either  due to  a
recollapse  (see  above) or vanishing of the  main  determinant.  The
fraction of such solutions increases with  initial $\phi$ decreasing.
For  some  value  of  initial  $\phi$  the solutions, nonsingular  in
future, disappear  completely  (apart  from the exceptional isotropic
case). This  critical  value depends on generalized
Hubble parameter $h$
as it is plotted in Fig.\,\ref{pminmax}. The transition from the
regime with only
nonsingular  future  solutions  to  one  with  only  singular  future
solutions  is rather  sharp  according to our  results  which can  be
easily seen  from Fig.\,\ref{pminmax}. The universe  with $\phi <
\phi_0$ (for  a
given $h$) and an arbitrary small nonzero initial  anisotropy can not
leave a  high-curvature regime and  falls into a singularity, so such
initial conditions are not  suitable  for describing the evolution of
the our Universe.

The other important dynamical feature  that  was  found via numerical
integrations is related with past asymptotic.  Though both described
types of singularity are valid for various initial data sets, all the
initial  data  which   lead   to  nonsingular  future  behavior  have
determinant singularity in the past (again, we ignore the exceptional
isotropic solution). In  particular,  for large initial $\phi$ the
only possible past singularity is the determinant singularity
(Fig.\,\ref{phi0}).

To obtain this result we  investigated  a rather wide set of  initial
conditions. However, the full set  of  allowed  initial conditions is
noncompact and,  hence, we have no  proof of this  statement in
the strict sense. Hence, we can put forward {\bf a  conjecture:} {\it
the past singularity for nonsingular in future Bianchi I cosmological
model with the second  order  curvature corrections is of determinant
type.} This proposal is analogous to the result obtained in \cite{e5}
for Schwarzschild  black  hole,  where  the  asymptotically Minkowski
initial data at  infinity lead to  a determinant singularity  at  the
finite values of the radial coordinate.

\section*{Acknowledgments}
This  work  was  supported  via 'Universities of  Russia,  Fundamental
Investigation' grant no 990777 and,  partially  supported  by  Russian
Foundation   for   Basic  Research  via  grants  no  99-02-16224   and
00-15-96699. Authors are grateful to H.-J.Schmidt for comments.

\onecolumn

\end{document}